\documentclass[superscriptaddress,amsmath,amssymb,aps,prl,twocolumn,floatfix]{revtex4-2}

\usepackage[utf8]{inputenc}
\usepackage{booktabs}
\usepackage{ifthen}
\usepackage[colorlinks, linkcolor=myblue, citecolor=myblue, urlcolor=myblue, breaklinks]{hyperref}
\usepackage{tikz}
\usepackage{siunitx}
\usepackage{rotating}

\definecolor{myblue}{RGB}{54,93,201}

\usepackage{bbm}
\usepackage{amsmath,amsfonts,amssymb}
\usepackage{cleveref}

\begin{document}

\title{A Network Approach to Atomic Spectra}

\author{David Wellnitz}
\thanks{These authors contributed equally.}
\affiliation{ISIS (UMR 7006) and IPCMS (UMR 7504), University of Strasbourg and CNRS, and icFRC, 67000 Strasbourg, France}
\affiliation{Physikalisches Institut, Universität Heidelberg, Im Neuenheimer Feld 226, 69120 Heidelberg, Germany}

\author{Armin Keki\'{c}}
\thanks{These authors contributed equally.}
\affiliation{Physikalisches Institut, Universität Heidelberg, Im Neuenheimer Feld 226, 69120 Heidelberg, Germany}
\affiliation{\'{E}cole Normale Sup\'{e}rieure, 75005 Paris, France}

\author{Julian Heiss}
\affiliation{Physikalisches Institut, Universität Heidelberg, Im Neuenheimer Feld 226, 69120 Heidelberg, Germany}

\author{Michael Gertz}
\affiliation{Institute of Computer Science, Universität Heidelberg, 69120 Heidelberg, Germany}

\author{Matthias Weidem\"uller}
\thanks{weidemueller@uni-heidelberg.de}
\affiliation{Physikalisches Institut, Universität Heidelberg, Im Neuenheimer Feld 226, 69120 Heidelberg, Germany}

\author{Andreas Spitz}
\thanks{andreas.spitz@uni-konstanz.de}
\affiliation{Department of Computer and Information Science, University of Konstanz, 78464 Konstanz, Germany}

\date{\today}

\begin{abstract}
Network science provides a universal framework for modeling complex systems, contrasting the reductionist approach generally adopted in physics. In a prototypical study, we utilize network models created from spectroscopic data of atoms to predict microscopic properties of the underlying physical system. For simple atoms such as helium, an a posteriori inspection of spectroscopic network communities reveals the emergence of quantum numbers and symmetries. For more complex atoms such as thorium, finer network hierarchies suggest additional microscopic symmetries or configurations. Link prediction yields a quantitative ranking of yet unknown atomic transitions, offering opportunities to discover new spectral lines in a well-controlled manner. Our work promotes a genuine bi-directional exchange of methodology between network science and physics, and presents new perspectives for the study of atomic spectra.
\end{abstract}

\maketitle

Network science~\cite{Newman2010, Barabasi2016} promotes a holistic perspective for the study of complex systems, allowing for a deeper understanding of their structure and dynamical behavior~\cite{Albert2000error} and aiding in identifying patterns in the interactions of the system’s components. Applications include social~\cite{granovetter1973strength,newman2001structure}, communication~\cite{BRODER2000309,faloutsos1999power}, ecological~\cite{Montoya2006,proulx2005network}, physiological~\cite{jeong2000large,rual2005towards}, and epidemiological~\cite{Keeling2005Jun,Meyers2012Jan} networks. In contrast, complex physical systems such as atoms, molecules, materials, or even the universe are generally approached from a microscopic, genuinely reductionist point of view, starting from fundamental sets of equations. With increasing complexity, computing the solutions to these equations represents a compelling, if not insurmountable challenge. As a consequence, the study of complex physical systems stands to benefit from the application of network science~\cite{Valdez2017}. Furthermore, complex physical systems with known ground truth can serve as a complementary paradigm for testing algorithms in network science, in contrast to commonly used social benchmarks with unknown ground truth~\cite{Zachary1977}.

In order to better understand a network, it is often promising to find patterns underlying the network's structure. In social networks, these patterns often take the form of communities with strong links within each community~\cite{simmel1950sociology,granovetter1973strength}, and many methods have been devised to identify such communities~\cite{Fortunato2010}. More generally, communities can be formalized in the framework of stochastic block models, which are defined according to statistically significant patterns in the connectivity matrix~\cite{holland1983stochastic,Peixoto2014hierarchical,Peixoto2019book}. Based on such knowledge about the network structure, it is also possible to make predictions about the network, such as its evolution or missing links~\cite{martinez2016survey,liben2007link}.

Here, we investigate the potential of network science to make predictions about atomic spectra as prototypical complex physical systems. We start by introducing a natural mapping from spectral data to networks. We then proceed to identify communities in these networks. Surprisingly, we find that these communities relate to the underlying quantum mechanical structure of the atoms, which we explain by selection rules. For the thorium II spectrum, we find additional patterns in the network, which are energetically correlated. Finally, we show how link prediction can be applied to predict yet unknown atomic transitions with great accuracy.

\begin{figure*}
	\centering
\includegraphics[width=0.8\textwidth]{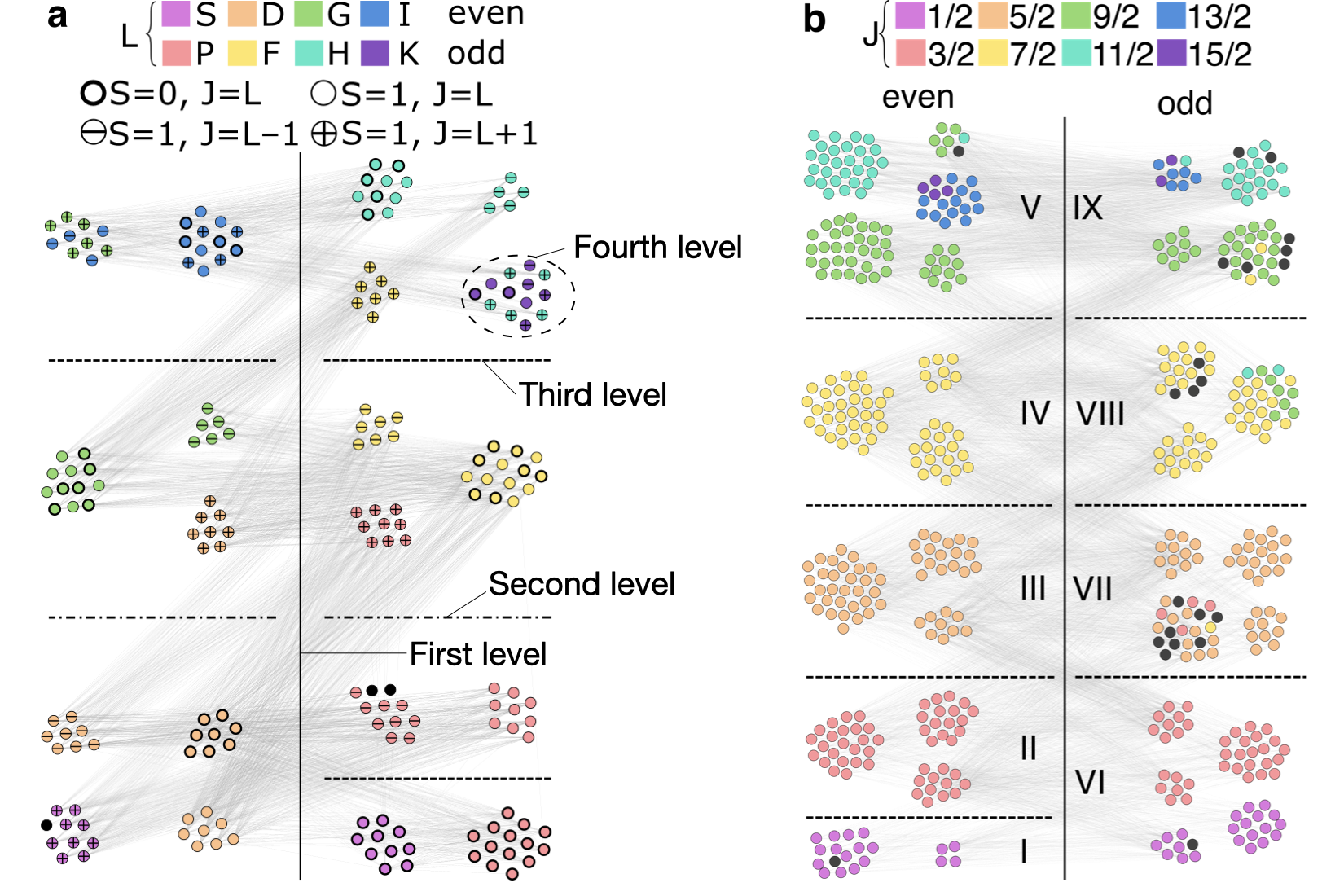}
\caption{Community structure of \textbf{a}, the helium I, and \textbf{b}, the thorium II network. The continuous vertical line denotes the highest hierarchy level, while the horizontal lines indicate intermediate hierarchy levels (for helium I, the dash-dotted lines indicate a higher, the dashed lines a lower intermediate level). Spatial arrangement marks the lowest hierarchy level.  Intermediate-level communities of thorium II are labeled with roman numerals.
The colors and symbols encode electronic quantum numbers such as the orbital angular momentum $L$, the spin $S$, and the total angular momentum $J$. The black nodes correspond to doubly excited states (helium I) and states with unassigned total angular momentum (thorium II), respectively.
}
	\label{fig:com}
\end{figure*}

\textit{Spectroscopic Networks --- } The spectroscopic data of an atom are a set of transitions between its internal energy levels (states) that characterize the atom's structure~\cite{svanberg2012atomic}. This spectroscopic data can be naturally viewed as a network by identifying energy levels with the nodes (vertices) of the network, and transitions with the links (edges) between nodes.
Such a mapping has been used previously to describe molecular spectroscopic data~\cite{Csaszar2016}, improve their accuracy and assignment~\cite{Csaszar2016,tobias2020spectroscopic}, identify errors~\cite{arendas2020bridges,Tobias2021May}, and design efficient measurements~\cite{tobias2020spectroscopic,arendas2020bridges}. Spectroscopic data can be obtained either from a solution of the Schr\"odinger equation for small systems such as hydrogen or helium, or from the empirical observation of transitions, as compiled, e.g., in the NIST atomic spectra database~\cite{Kramida2018}. In the following, helium I (neutral helium) serves as an example of a numerically solvable system, while we investigate thorium II (singly ionized thorium) as a more complex system without adequate \emph{ab initio} solutions.

The resulting spectroscopic networks of helium I and thorium II are not scale-free (see Appendix), in contrast to many other network types, including molecular spectroscopic networks, which follow scale-free degree distributions~\cite{Csaszar2016,barabasi1999emergence}. In fact, one may argue that spectroscopic networks cannot be accurately described by typical emergence mechanisms behind complex networks, such as preferential attachment~\cite{barabasi1999emergence}, since their structure is determined by microscopic laws and symmetries~\cite{Bransden2003} .

\textit{Community Detection --- } To identify patterns with predictive power in spectroscopic networks, we define communities by grouping nodes according to their link structure~\cite{Fortunato2010}. While community detection using modularity maximization has been applied to spectroscopic networks before, it did not reveal the full underlying community structure~\cite{Csaszar2016}. Unlike social networks, whose community structure arises from triadic closure~\cite{granovetter1973strength} and thus supports community detection based on modularity maximization~\cite{Girvan2002}, spectroscopic networks are actually characterized by an anti-community structure imposed by the parity symmetry that can be detected by modularity minimization~\cite{Lackner2018}. In order to avoid bias, we employ an agnostic approach that detects communities and anti-communities equally well: the nested stochastic blockmodel (NSBM), which generates a hierarchical partitioning of nodes with similar connections~\cite{Peixoto2019book}.

In the helium I network, the community detection algorithm reveals four different hierarchy levels displayed in Fig.~\ref{fig:com}a. On the first (coarsest) hierarchy level, an almost exact bipartite structure is identified with only $\sim 5\%$ of the links connecting nodes within each of the two communities. This reflects the above-mentioned parity symmetry, resulting in corresponding selection rules for the spectroscopic lines~\cite{Csaszar2016}. On the next hierarchy levels, both communities are divided further into clusters that reflect the underlying quantum mechanical structure of the states. In particular, the second and third hierarchy level correspond to a successively finer separation into ranges of total angular momentum $J$, except for the \textsuperscript{1}S\textsubscript{0} and \textsuperscript{1}P\textsubscript{1} states. For example, for the even parity states, we have one community with $0 < J \leq 2$, one with $2 < J \leq 4$, and one with $4 < J \leq 7$. The final, fourth hierarchy level groups states along their quantum numbers for spin, orbital and total angular momentum as displayed in Fig.~\ref{fig:com}a.

This correspondence of community structure to the underlying microscopic physical principles results from selection rules, which classify transitions as allowed or forbidden depending on the quantum numbers~\cite{Bransden2003, Morton2006}. Although generally a transition between any pair of states is possible, allowed transitions are typically much stronger than forbidden transitions, and thus more likely to be included in the spectroscopic network. In fact, we can assume that the identification of these quantum numbers is mostly limited by the finite resolution of the NSBM, i.e.~by a minimal required number of nodes per community~\cite{Peixoto2019book}, because a mixing of states with different quantum numbers only occurs for high angular momentum, for which fewer states are registered in the NIST database.

As an example for the application of community detection to a more complex system, we focus on the spectrum of thorium II, for which a large body of spectroscopic data is available, and which is of practical interest for the construction of a nuclear clock~\cite{Peik2003,Porsev2010}. Due to its complex electronic structure as an actinide with three valence electrons, many symmetries are broken, rendering an \emph{ab initio} solution of the Schr\"odinger equation impossible. In fact, not even basic quantum numbers such as the total angular momentum can be assigned to every state~\cite{Safronova2014, Redman2014}.

Three hierarchy levels are identified by the NSBM as shown in Fig.~\ref{fig:com}b. We find that the first hierarchy level again yields a perfectly bipartite separation of nodes, reflecting parity as a good quantum number. The second hierarchy level essentially corresponds to the states with a given total angular momentum quantum number $J$. Nodes indicated in black in Fig.~\ref{fig:com}b had no \emph{a priori} assignment in the database so far. Based on the underlying community structure, a corresponding $J$ value might be associated \emph{a posteriori}~\cite{wellnitzma}.


\begin{figure}
	\centering
	\includegraphics[width=\columnwidth]{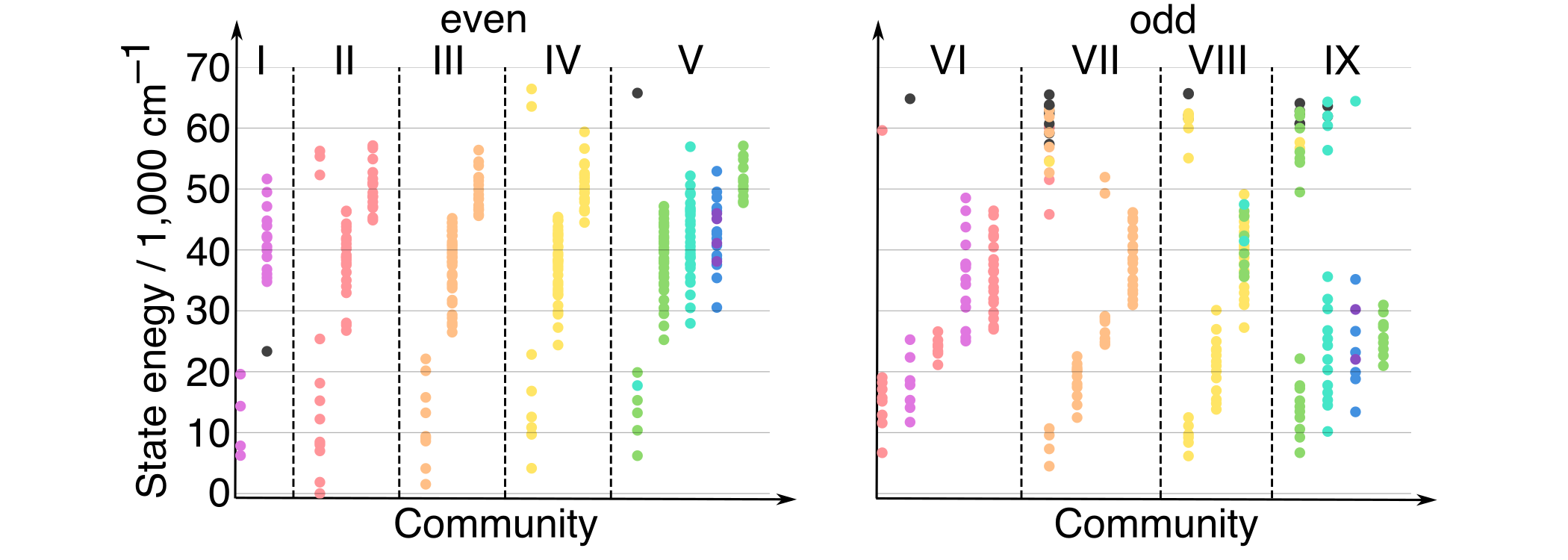}
	\caption{Energies of thorium II states sorted by their respective community. Each point represents one node in the network (color coding and hierarchy assignment as in Fig.~\ref{fig:com}). Communities are sorted by increasing average energy within the first hierarchy level.
}
	\label{fig:energies}
\end{figure}

For the third hierarchy level however, we find no obvious quantum mechanical correspondence, except for a finer resolution of few high angular momentum states $J>9/2$. To gain further insight into the microscopic interpretation of this hierarchy level, we arranged the thorium II nodes according to the energies of the corresponding states (see Fig.~\ref{fig:energies}). Interestingly, the different communities form clusters within a contiguous energy range, indicating that states represented by a community share common physical properties. One might speculate that the orbital structure of these states bares some similarities as further supported by correlations between communities and occupied electron orbitals (see Appendix). Therefore, these communities might indicate hidden symmetries that could be exploited for applying advanced methods in electronic structure calculations such as Configuration Interaction~\cite{Safronova2014,sherrill1999configuration}.


\begin{figure}
	\centering
	\includegraphics[width=\columnwidth]{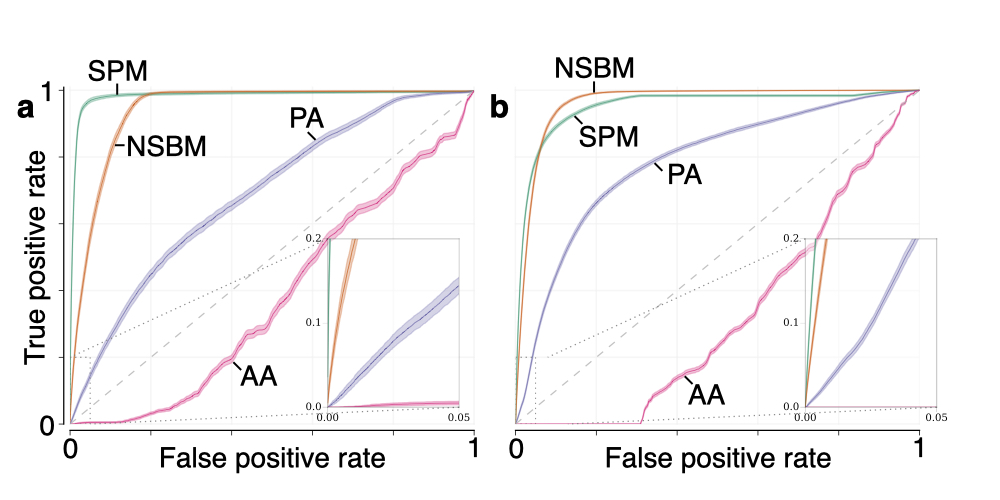}
	\caption{
Receiver operating characteristic (ROC) for link prediction on \textbf{a} the helium I and \textbf{b} the thorium II network with the structural perturbation method (SPM, green), the nested stochastic block model (NSBM, orange), the preferential attachment index (PA, purple), and Adamic Adar index (AA, pink). For each spectroscopic network, 10\% of links are removed uniformly at random and used as a ground truth for subsequent prediction (dropout method)~\cite{Fawcett2006}. The ROC is constructed by iterating over the computed ordered list of predictions. Each incorrect prediction moves one step to the right, and each correct prediction one step up. The shaded area indicates the standard error of the different realizations.
}
	\label{fig:ROC}
\end{figure}%

\textit{Link prediction --- } Link prediction is a network tool to infer missing links from the network structure~\cite{martinez2016survey}, often by emulating a natural densification process of implicit community structure, such as triadic closure in social networks~\cite{simmel1950sociology,granovetter1973strength}. A link prediction algorithm ranks all absent links according to the estimated likelihood that they should be included. In the following, we demonstrate how such methods can be utilized to predict previously unobserved atomic transitions without having to rely on any quantum mechanical model.
From the available link prediction algorithms, we select four representative methods. The Adamic-Adar (AA) index~\cite{Adamic2003} and the preferential attachment (PA) index~\cite{liben2007link} are standard methods designed for the prediction of evolving friendships in social networks. The nested stochastic block that we used for community detection also supports link prediction~\cite{Peixoto2019book}. Finally, the structural perturbation method (SPM) is a state-of-the-art method that derives link likelihoods from perturbations of the network's adjacency matrix~\cite{Lu2015a}.

For a quantitative evaluation, we simulate missing links through dropout, that is, by removing a randomly selected subset of links from the network and subsequently attempting to recover them. In Fig.~\ref{fig:ROC}, we show the receiver operating characteristic (ROC) curve~\cite{Fawcett2006} of the results on the helium I and thorium II networks. PA and especially AA perform poorly due to the anti-community structure of the spectroscopic networks, indicating that traditional link prediction methods used, e.g., in social networks, are not suited.
In contrast, SPM and NSBM perform very well for the networks of both atoms. While the SPM shows slightly better results when aiming for the discovery of new transitions, the NSBM provides better fidelity in retrieving the complete set of missing lines. Based on this excellent performance in the dropout approach, we applied these methods to predict yet unobserved transitions, which are listed in tables S1 to S4. Such predictions may guide future high precision spectroscopic measurements looking for yet unmeasured transitions.

\textit{Conclusion --- } Our results indicate that the holistic character of network science offers a promising path to revealing additional insights into the structure of complex physical systems, similar to previous suggestion for Ising and Hubbard models~\cite{Valdez2017}. Furthermore, complex physical systems such as atomic spectra, for which the solutions of the microscopic equations serve as a ground truth, can be considered a complementary paradigm for testing algorithmic methods in network science, in contrast to, e.g., commonly used social benchmarks~\cite{Zachary1977}, for which no \emph{ab initio} ground truth is known. In this work, we focus on helium I and thorium II for conciseness, but we have already extended community detection to iron, and performed link prediction for carbon and iron yielding similar results (see Appendix). The approach presented here can readily be generalized to other discrete spectra, such as molecular or nuclear spectra~\cite{Csaszar2016,hitran2016}. Promising future applications of concepts from network science to spectroscopy include the derivation of transition strengths from weighted networks~\cite{Zhao2015}, or the prediction of yet unobserved atomic states by emerging node prediction methods~\cite{wellnitzma,kim2011network}, providing a guideline to experimental and theoretical studies of atomic structure.

\section*{Acknowledgements}
The authors thank Sebastian Lackner, Kathinka Gerlinger, Guido Pupillo and Johannes Schachenmayer for discussions and comments. This work is supported in part by the Deutsche Forschungsgemeinschaft (DFG, German Research Foundation) under Germany’s Excellence Strategy EXC2181/1-390900948 (the Heidelberg STRUCTURES Excellence Cluster) and the Heidelberg Center for Quantum Dynamics. This work was supported in part by the LabEx NIE (``Nanostructures in Interaction with their Environment'') under contract ANR-11-LABX0058 NIE as part of the ``Investments for the future program'', and the Agence Nationale de la Recherche (Grant ANR-17-EURE-0024 EUR QMat). Armin Keki\'c would like to thank the German Academic foundation (Studienstiftung des deutschen Volkes) for their support.

\bibliography{networks}

\newpage

\appendix

\begin{widetext}

\section*{Appendix}

\subsection*{Data Selection}
We analyzed two datasets from the NIST atomic spectra database~\cite{Kramida2018}: (i) For helium I, we used the given 193 energy levels and 2300 transitions, and (ii) for thorium II, we used the given 516 energy levels and 6502 transitions. Each dataset was cast into a network as described in the first paragraph of the paper. For the subsequent analysis, we used the software libraries \texttt{networkx}~\cite{Hagberg2008networkx} and \texttt{graph-tool}~\cite{Peixoto2014graphtool}.

Many link prediction algorithms are not designed to predict the connections between separate connected components, such as the Adamic-Adar index~\cite{Adamic2003} and the structural perturbation method~\cite{Lu2015a}. Therefore, we limit our analysis to only the largest connected component of each network. This component contains 191 nodes and 2299 links for helium I (i.e.~we remove a single disconnected component of two nodes and one link), and all nodes and links for thorium II.

\subsection*{Link Prediction Dropout} 
In the dropout procedure used to evaluate the link prediction, the network can become disconnected. In this case, the link prediction was performed on the largest connected component. Since less than one node is removed from either network on average, the impact on the results is negligible.

\subsection*{Community Detection}
We used a nested stochastic blockmodel approach to detect communities~\cite{Peixoto2019book}. This model assigns an entropy to each possible hierarchical partition of the network. The entropy quantifies the amount of information that the partition contains about the network structure, such that minimal entropy corresponds to the most likely partition. A Markov chain Monte Carlo algorithm implementation from \texttt{graph-tool} was used to find the partition with minimal entropy. This algorithm was executed 100 times, and the lowest entropy fit was plotted using Gephi~\cite{bastian2009gephi} in Fig.~\ref{fig:com}. The other runs yielded similar results, as shown in the Fig.~S2, and led to the same conclusions. Details on the implementation and the different communities are given in the supplementary material.

\subsection*{NSBM Link Prediction}
The probability of each undetected link $l$ is given by $p(l) \propto \sum_\mathbf{b} p(l|\textbf{b}) P(\mathbf{b})$ with $P(\mathbf{b}) = \exp(-S(\textbf{b}))$, where the $\mathbf b$ is a network partition, $p(l|\mathbf b$) is the probability of link $l$ for a given partition $\mathbf b$, and $S(\mathbf b)$ is the entropy of $\mathbf b$~\cite{Peixoto2019book}. We approximated the sum by sampling 100 partitions with a Markov chain Monte Carlo algorithm (implemented in Ref.~\cite{Peixoto2014graphtool}) that returns a set of likely partitions, starting from a low entropy partition. For each partition, we computed its entropy $S(\mathbf b)$, and took the sum over all different partitions with $S(\mathbf b) - S_\mathrm{min} \leq 1$, where $S_\mathrm{min}$ is the smallest entropy. We confirmed convergence for the chosen number of samples and entropy cutoff.

\subsection*{Degree Distributions}

\begin{center}
\includegraphics[width=0.8\textwidth]{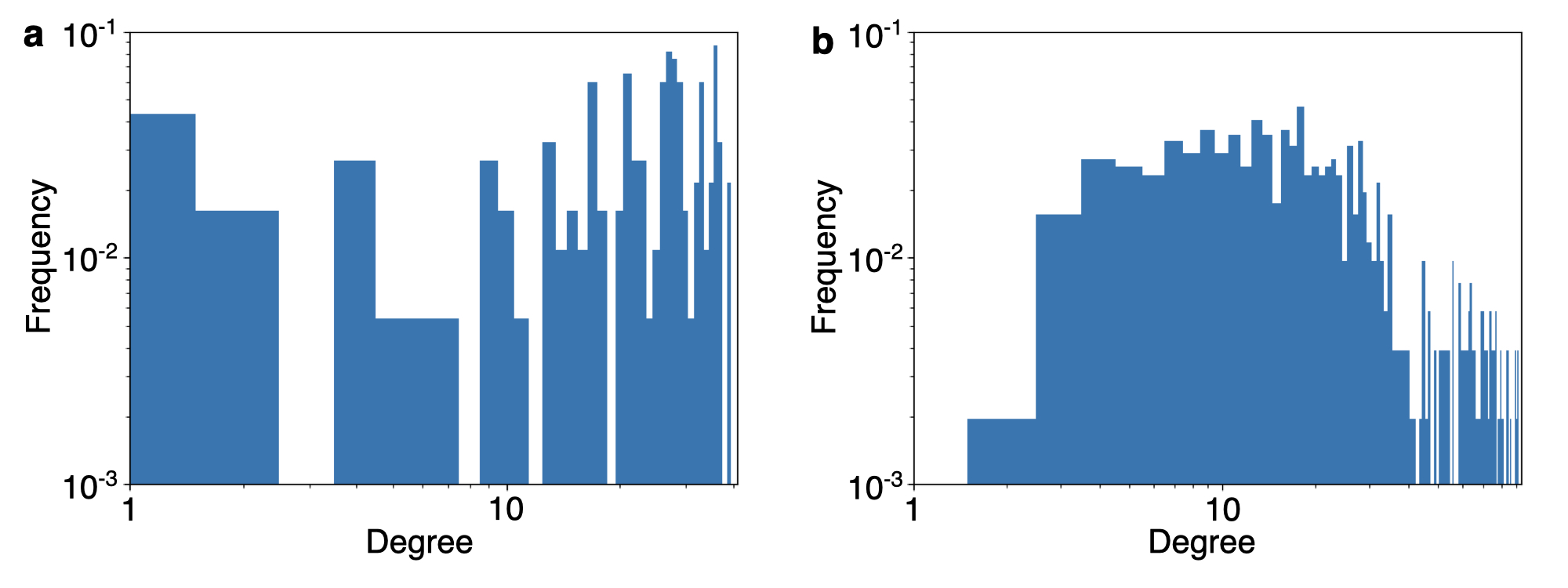}
\end{center}

\noindent\textbf{Fig.~S1: Degree distributions.} The relative frequency of the number of links for each node for \textbf{a}, the helium I network, and \textbf{b}, the thorium II network. A scale-free model would correspond to a power law type distribution, which is not found here.

\subsection*{Comparison of Partitions}

\begin{center}
\includegraphics[width=\textwidth]{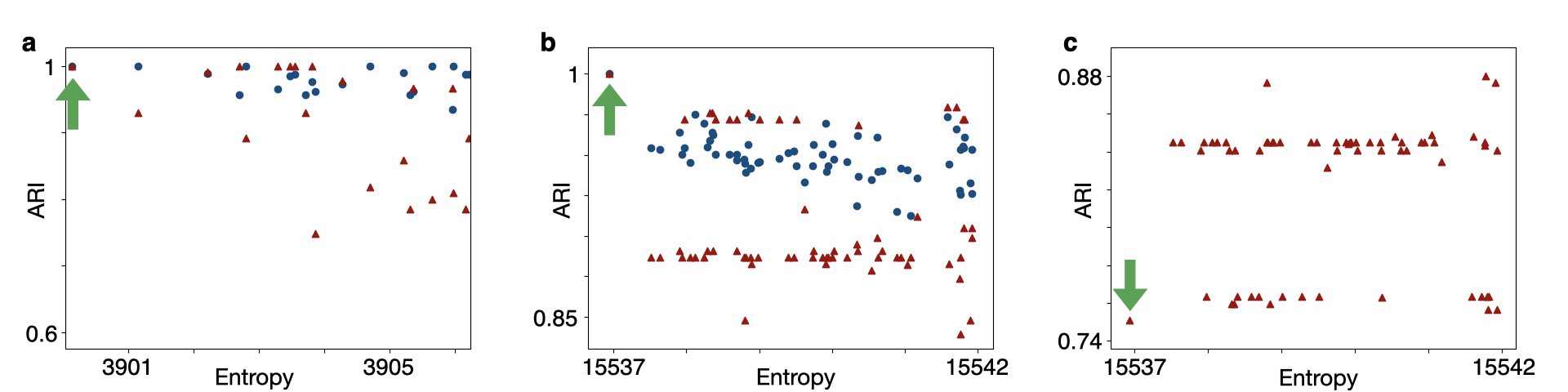}
\end{center}

\noindent\textbf{Fig.~S2: Comparison of different possible partitions.} The Adjusted Rand Index (ARI) measures the similarity of different partitions, with zero corresponding to random correlations, and one corresponding to identical partitions~\cite{Hubert1985}. For both helium I and thorium II, multiple low entropy partitions found by the community detection algorithm are considered. The ARI is plotted against their entropy, which measures the partition quality in the NSBM framework~\cite{Peixoto2019book}. In \textbf a, the different partitions of the helium I network are compared to the partition shown in Fig.~1 of the main text, indicated by the green arrow. Blue dots correspond to the lowest hierarchy level, red triangles to the first intermediate hierarchy level. Panel \textbf b shows the results for thorium II (same formatting as \textbf a). Panel \textbf c shows a comparison to communities formed by equal $J$ and parity, where nodes with unknown $J$ are ignored for thorium II. All identified partitions show an even better correspondence between communities and quantum numbers that the one shown in Fig.~1b.

\subsection*{Relation to Quantum Numbers}

\begin{center}
\includegraphics[width=0.8\textwidth]{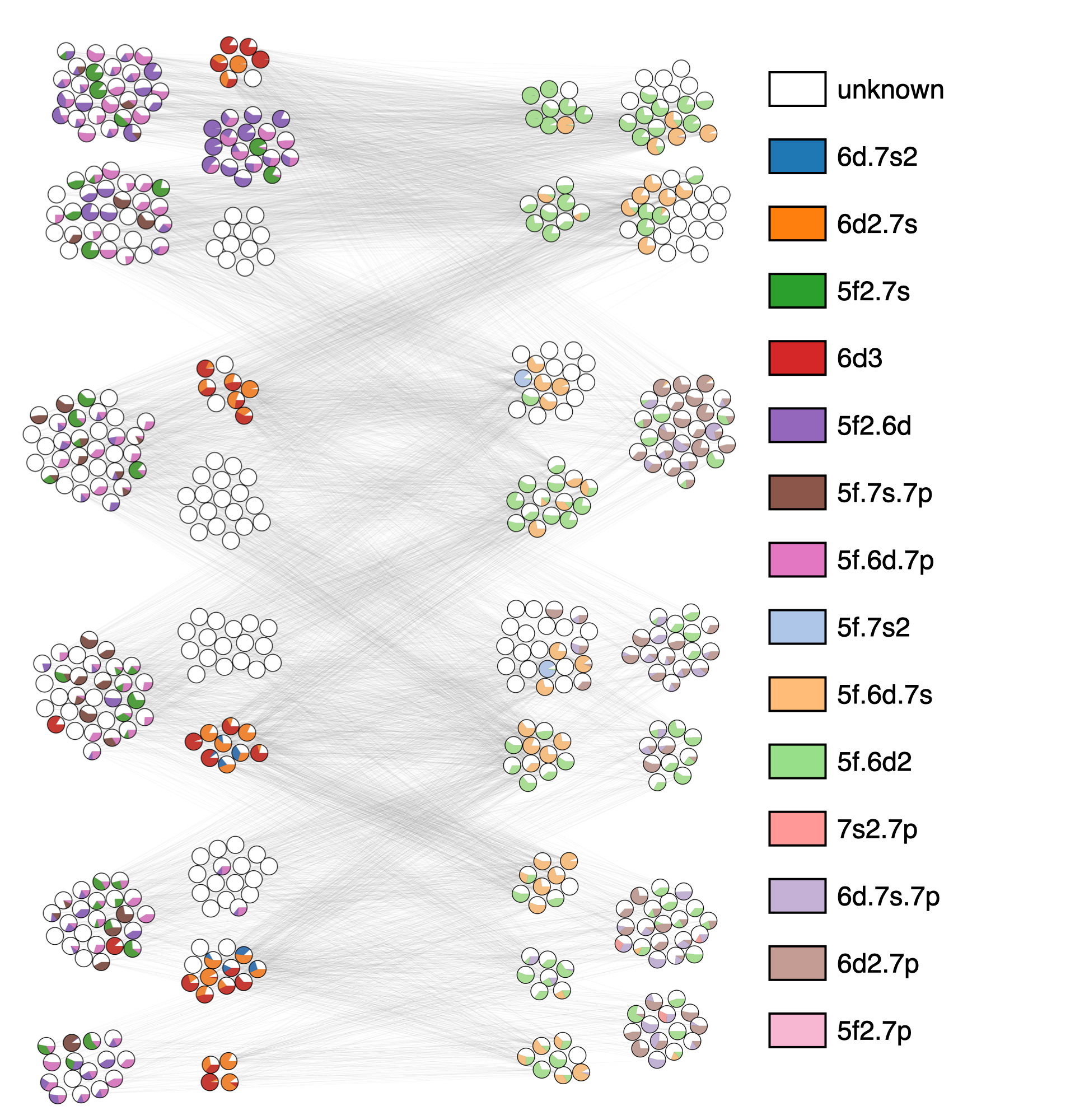}
\end{center}

\noindent\textbf{Fig.~S3: Correlation between communities and occupied electron orbitals.} The node positions correspond to the positions given in Fig.~1 of the main text. Each node is a pie chart of the contributing orbital structure. The two dominant configurations are taken from the NIST database and the electron orbitals are extracted from these configurations. For each state, a pie chart of the orbitals with their corresponding quantum mechanical occupation probabilities is generated. Unassigned probability is left white, and can correspond any configuration. For even parity states, we find a clear separation between states with and without contributions from the 5$f$ orbital at the lowest hierarchy level. For odd parity states, we find no single important orbital, but the $6d^27p$ and the $5f6d7s$ orbitals are sorted into different communities.

\subsection*{Iron Results}

\begin{center}
\includegraphics[width=0.8\textwidth]{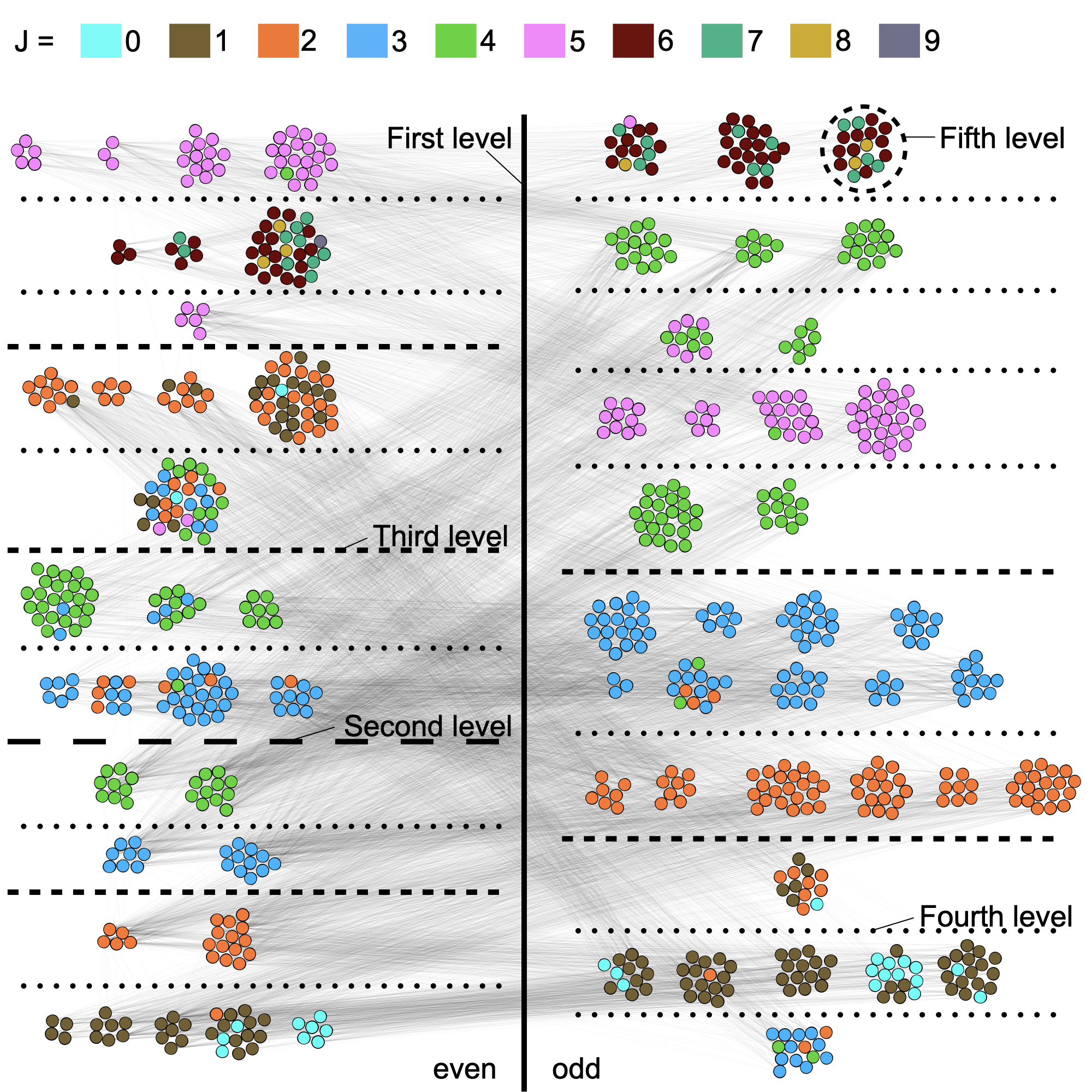}
\end{center}

\noindent\textbf{Fig.~S4: Communities of the Fe I spectrum.} Five hierarchy levels identified by the nested stochastic blockmodel for the network generated from neutral iron are depicted. The network is made up of 846 nodes and 9897 links. The hierarchy levels are indicated by different line types and spacial arrangement, while the total angular momentum $J$ is encoded by color, analogous to Fig.~\ref{fig:com}b in the main text. We find perfect correspondence between the first hierarchy level and the parity quantum number of the corresponding states, as indicated by the even and odd label. Furthermore, we find that on the fourth level, states whose nodes are in the same community are likely to share the same total angular momentum quantum number, analogous to our analysis of thorium II in the main text. See also Ref.~\cite{wellnitzma} for a more detailed discussion.

\begin{center}
\includegraphics[width=0.7\textwidth]{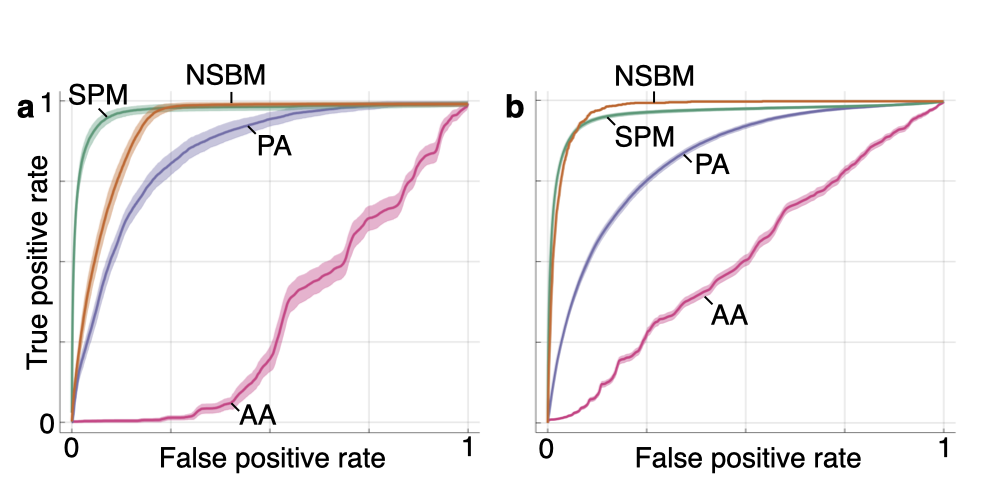}
\end{center}

\noindent\textbf{Fig.~S5: Link prediction for carbon and iron.} The receiver operating characteristics for link prediction in the network generated from \textbf{a} neutral carbon and \textbf{b} neutral iron with the methods discussed in the paper are shown, analogous to Fig.~\ref{fig:ROC} in the main text. The carbon network contains 180 nodes and 1377 links. The Adamic Adar index performs worst in all cases, followed by preferential attachment. The structural perturbation method performs best for the top ranked predictions in all cases, whereas the nested stochastic blockmodel performs is more accurate for low ranked predictions, as discussed in the paper. The prediction was run after a random dropout of 10\% of the links. Due to computational cost, only one run was performed with the NSBM on iron. In all other cases the mean and standard deviation of the mean computed from 100 different dropouts are displayed. See also Ref.~\cite{heissma} for more  methods and a more detailed discussion.

\subsection*{Link Prediction Results}

\begin{center}
	
\fontfamily{phv}\selectfont
\fontsize{7}{8}\selectfont

\begin{tabular}{lrrllll}
\toprule
Score     &       Energy 1 &       Energy 2 & Term 1 & Term 2 &  Conf 1 &  Conf 2 \\
\midrule
13.304399 &  196955.226641 &  196955.946185 &    3D2 &    3F2 &   1s.9d &   1s.9f \\
9.977080  &  195262.726142 &  197213.352292 &    3G3 &    3F4 &   1s.6g &  1s.10f \\
8.910412  &  196071.415767 &  197213.420506 &    3H4 &    3G5 &   1s.7h &  1s.10g \\
8.252403  &  197213.437274 &  197213.420506 &    1H5 &    3G5 &  1s.10h &  1s.10g \\
7.463494  &  195262.726142 &  195262.795043 &    3G3 &    3H4 &   1s.6g &   1s.6h \\
4.935619  &  196071.371284 &  196071.415767 &    1G4 &    3H4 &   1s.7g &   1s.7h \\
4.935619  &  196071.368738 &  196071.415767 &    3G4 &    3H4 &   1s.7g &   1s.7h \\
4.715659  &  196956.061082 &  196956.067880 &    1H5 &    3I5 &   1s.9h &   1s.9i \\
4.438541  &  197213.442433 &  197213.436916 &    1I6 &    3H6 &  1s.10i &  1s.10h \\
4.370201  &  197213.420506 &  197213.352292 &    3G5 &    3F4 &  1s.10g &  1s.10f \\
\bottomrule
\end{tabular}

\end{center}

\noindent\textbf{Table S1: Predicted transitions by the NSBM for helium I.} The top ten predictions from the full network are given, ranked by the prediction score. The states are identified by their energies in cm$^{-1}$. Additionally, the term and electron configuration for both states is given. All predicted transitions are dipole allowed.

\begin{center}
	
\fontfamily{phv}\selectfont
\fontsize{7}{8}\selectfont

\begin{tabular}{lrrllll}
\toprule
Score    &       Energy 1 &       Energy 2 & Term 1 & Term 2 &  Conf 1 &  Conf 2 \\
\midrule
0.445071 &  195262.795750 &  195262.724684 &    1H5 &    3G5 &   1s.6h &   1s.6g \\
0.430666 &  191451.897461 &  191444.482131 &    1F3 &    3D2 &   1s.4f &   1s.4d \\
0.421943 &  196069.676490 &  196071.178730 &    3D1 &    3F2 &   1s.7d &   1s.7f \\
0.416554 &  195262.794095 &  195262.724684 &    3H6 &    3G5 &   1s.6h &   1s.6g \\
0.411815 &  193921.125753 &  193917.151287 &    3F2 &    3D3 &   1s.5f &   1s.5d \\
0.397299 &  193921.617719 &  193921.121343 &    3G5 &    3F4 &   1s.5g &   1s.5f \\
0.384420 &  196069.673050 &  196071.178730 &    3D2 &    3F2 &   1s.7d &   1s.7f \\
0.377473 &  196595.062365 &  196596.080442 &    3D2 &    3F2 &   1s.8d &   1s.8f \\
0.373079 &  196955.946185 &  196955.228258 &    3F2 &    3D1 &   1s.9f &   1s.9d \\
0.367857 &  195262.724684 &  195262.793051 &    3G5 &    3H5 &   1s.6g &   1s.6h \\
\bottomrule
\end{tabular}

\end{center}

\noindent\textbf{Table S2: Top 10 predicted transitions by the SPM for helium I.} The top ten predictions from the full network are given, ranked by the prediction score. The states are identified by their energies in cm$^{-1}$. Additionally, the terms and electron configurations of both states are given. All predicted transitions are dipole allowed.

\begin{center}
	
\fontfamily{phv}\selectfont
\fontsize{7}{8}\selectfont

\begin{tabular}{lrrll}
\toprule
Score    &     Energy 1 &     Energy 2 &   J 1 &  J 2 \\
\midrule
7.641832 &   1859.93843 &  40472.45100 &   3/2 &  3/2 \\
7.587684 &  40472.45100 &  22106.43260 &   3/2 &  5/2 \\
4.870402 &  17121.62038 &  33730.93510 &   3/2 &  5/2 \\
4.701044 &   7828.56081 &  40472.45100 &   1/2 &  3/2 \\
4.661393 &  44807.93520 &  10673.13832 &   7/2 &  5/2 \\
4.618527 &  44807.93520 &   8378.85915 &   7/2 &  7/2 \\
4.613255 &   4146.57708 &  40706.81310 &   7/2 &  7/2 \\
4.552611 &   1521.89632 &  40472.45100 &   5/2 &  3/2 \\
4.542269 &  15453.03596 &  33730.93510 &   7/2 &  5/2 \\
4.515303 &  15710.84204 &  42336.82990 &   3/2 &  5/2 \\
\bottomrule
\end{tabular}

\end{center}

\noindent\textbf{Table S3: Predicted transitions by the NSBM for thorium II.} The top ten predictions from the full network are given, ranked by the prediction score. The states are identified by their energies in cm$^{-1}$. Additionally, the total angular momenta of both states is given. All predicted transitions are dipole allowed.

\begin{center}
	
\fontfamily{phv}\selectfont
\fontsize{7}{8}\selectfont

\begin{tabular}{lrrll}
\toprule
Score    & Energy 1 &     Energy 2 &   J 1 &   J 2 \\
\midrule
0.200714 &  21297.41708 &   4146.57708 &   5/2 &   7/2 \\
0.188276 &  17460.62698 &   1521.89632 &   5/2 &   5/2 \\
0.178387 &  12570.49403 &  21682.74765 &   7/2 &   7/2 \\
0.176379 &  12902.37757 &  34543.55740 &   3/2 &   5/2 \\
0.172929 &   4146.57708 &  24982.44580 &   7/2 &   7/2 \\
0.170751 &  30452.72533 &   9202.26506 &   9/2 &   7/2 \\
0.169594 &   9585.40432 &  31259.29670 &   5/2 &   5/2 \\
0.168869 &   8460.35308 &  21131.79959 &   3/2 &   3/2 \\
0.166028 &  37063.30650 &  14275.57661 &   9/2 &   9/2 \\
0.164764 &   9720.29776 &  37787.87910 &   7/2 &   7/2 \\
\bottomrule
\end{tabular}

\end{center}

\noindent\textbf{Table S4: Top 10 predicted transitions by the SPM for thorium II.} The top ten predictions from the full network are given, ranked by the prediction score. The states are identified by their energies in cm$^{-1}$. Additionally, the total angular momenta of both states is given. All predicted transitions are dipole allowed.

\end{widetext}

\end{document}